\documentclass[aps,prl,showpacs,twocolumn,superscriptaddress]{revtex4}%
\usepackage{epsfig,dsfont,amssymb,amsmath,amsthm,amsfonts,amsbsy,mathrsfs}
\usepackage{graphicx}
\usepackage{graphicx}
\usepackage{color}
\usepackage{ulem}

\begin{document}

\title{Continuous Self Energy of Ions at the Dielectric Interface}

\author{Rui Wang}

\author{Zhen-Gang Wang}
\email {zgw@caltech.edu}
\affiliation{Division of Chemistry and Chemical Engineering,California Institute of Technology, Pasadena, CA 91125, USA}


\begin{abstract}

By treating both the short-range (solvation) and long-range (image
force) electrostatic forces as well as charge polarization induced
by these forces in a consistent manner, we obtain a simple theory
for the self energy of an ion that is continuous across the
interface.  Along with nonelectrostatic contributions, our theory
enables a unified description of ions on both sides of the
interface.  Using intrinsic parameters of the ions, we predict the
specific ion effect on the interfacial affinity of halogen anions at
the water/air interface, and the strong adsorption of hydrophobic
ions at the water/oil interface, in agreement with experiments and
atomistic simulations.

\end{abstract}


\maketitle

The interfacial activities of salt ions are of great importance in
physical chemistry, colloidal science and
biophysics\cite{Israelachvili}.  Many interfacial phenomena, such as
the surface tension of electrolyte solution\cite{Surfacetension},
salt effects on bubble coalescence\cite{Bubble} and effectiveness of
salts on the stability of proteins solutions and colloidal
suspensions\cite{Stability}, exhibit strong dependence on the
chemical identity of the ions.  Although this ``specific ion
effect'' has been known for over a century\cite{Hofmeister}, a
systematic, unified and predictive theory remains an outstanding
challenge.  Current theories are system dependent and require
adjustable parameters to force-fit experimental
data\cite{Ninhamrev,dispersion,shortrange,hydrophobic,Bier}.

A key factor that determines the ion distribution at the dielectric
interface and other interfacial properties is the self energy of a
single ion\cite{Wang1}.  The self energy consists of electrostatic
and nonelectrostatic contributions, such as cavity energy, hydration
and dispersion forces.  While the effects and the theoretical
treatments of these nonelectrostatic contributions are still
debatable \cite{Ninhamrev}, the constituent components in the
electrostatic self energy have become clear in recent
years\cite{Levin1,Levin2,Levin3}.  The  problem is then in the accurate
and consistent treatment of the electrostatic effects.  Such a
treatment is essential both because the electrostatic part is a
major component in the self energy of an ion, and because the
relative importance of the nonelectrostatic contributions can only
be evaluated when the electrostatic contribution is treated
accurately.

A major contribution in the electrostatic self energy is the image interaction, whose treatment was pioneered by Wagner, Onsager and Samaras
(WOS)\cite{Onsager}. The WOS theory predicts depletion of ions from
the water/air interface due to the image charge repulsion and qualitatively explains the
increase of surface tension with the salt concentration. However,
this theory fails to capture the initial decrease with salt
concentration in the surface tension known as the Jones-Ray
effect\cite{JonesRay}, and the systematic dependence on the identity
of the ions\cite{Surfacetension}.  A major weakness in the WOS theory
and its subsequent modifications is modeling the ion as a point
charge, which results in a discontinuous self energy across the
dielectric interface. The self energy diverges to positive infinity
on approaching the interface from the water side and to negative infinity on approaching from the air(oil) side.  To avoid this unrealistic behavior, the ion
distribution is artificially restricted to lie only in the water
phase, which makes the theory inapplicable to hydrophobic ions and
liquid-liquid interfaces.  This artificial cut-off also affects the
electrostatic potential gradient across the interface, which is
shown essential to the Jones-Ray effect\cite{Onuki}.

Another important effect is the finite polarizability of the ions.
Simulation by Jungwirth and Tobias\cite{Jungwirth} showed that the
polarizability of ions is a key contribution to their differential
affinity to the interface. Levin and coworkers
\cite{Levin1,Levin2,Levin3} developed a model of polarizable ions
near a dielectric interface that are able to explain several
interfacial properties of aqueous electrolyte solutions. In their
model, charge polarization in the ion is included to optimize the
short-range Born energy.  However, near a dielectric interface, the
long-range image force can be sufficiently strong to contribute to
charge polarization. Furthermore, their model does not account for
the image force on the air(oil) side of the interface, thus making
it difficult to extend the theory to hydrophobic ions and
liquid/liquid interfaces.

In this Letter, we present a unified theory that treats all the
electrostatic contributions: the Born solvation energy, the image
charge interaction, and ion polarizability in a single, consistent
framework. Along with the relevant nonelectrostatic contributions,
we apply our theory to air/water and liquid/liquid interfaces.

\textbf{Electrostatic Self Energy} We consider a single ion in the
vicinity of a sharp interface, located at $z=0$, between two
semi-infinite regions ($\Re_1$ and $\Re_2$) with respective
dielectric constant $\varepsilon_{1}$ and $\varepsilon_{2}$
($\varepsilon_{1}>\varepsilon_{2}$). We take the elementary charge
$e$ as the unit of charge, and $kT$ as the unit of energy. The ion
is taken as a sphere of radius $a$ centered at ${\bf r}_{c}$, with
charge distribution $\rho ({\bf r}, {\bf r}_{c})$, which satisfies
$\int d {\bf r} \rho ({\bf r}, {\bf r}_{c}) =\nu_{\pm}$ with
$\nu_{\pm}$ the valency of the ion (``$+$'' for cation and ``$-$"
for anion). The ion is polarizable; therefore, the charge
distribution will be self-adjusted to the local dielectric
environment.

The electrostatic self energy $u_{el}$ can be written as two parts:
$u_{el}=u_{int}+u_{pol}$, where $u_{int}$ accounts for the sum of
the Coulomb interactions in the constituent charges on the ion and
$u_{pol}$ is the energy cost of charge polarization. $u_{int}$ is
given by:
\begin{equation}
u_{int}({\bf r}_c)=2\pi l_B \int d {\bf r} \int d {\bf r}' \rho
({\bf r}, {\bf r}_{c}) G({\bf r}, {\bf r}') \rho ({\bf r}', {\bf
r}_{c}) \label{eq2}
\end{equation}
where $l_B=e^2/4\pi \varepsilon_0 kT$ is the Bjerrum length in the
vacuum and $\varepsilon_0$ is the vacuum permitivity. $G({\bf r},
{\bf r}')$ is the Green's function: the electrostatic potential at
${\bf r}$ due to a unit point charge at ${\bf r}'$. It satisfies the
Poisson equation $-\nabla \cdot \left[ \varepsilon ({\bf r}) \nabla
G({\bf r}, {\bf r}') \right]= \delta({\bf r}-{\bf r}')$ . Depending
on whether ${\bf r}$ and ${\bf r}'$ are in the same region, $G({\bf
r}, {\bf r}')$ is given by
\begin{eqnarray}
G({\bf r}, {\bf r}')=\begin{cases}
\frac {1} {4\pi \varepsilon_{\alpha} \mid {\bf r}-{\bf r}' \mid} + \frac{ \Delta_{\alpha \beta}}{4\pi \varepsilon_{\alpha} \mid {\bf r}-{\bf r}^{\star} \mid} & {\bf r},{\bf r}' \in \Re_{\alpha}\\
\frac {1} { 2\pi( \varepsilon_{\alpha}+ \varepsilon_{\beta}) \mid {\bf r}-{\bf r}' \mid}  & {\bf r} \in \Re_{\beta},{\bf r}' \in \Re_{\alpha}\\
\end{cases}  \label{eq3}
\end{eqnarray}
$\alpha$ and $\beta$ can be either 1 or 2, and $\Delta_{\alpha
\beta}=(\varepsilon_{\alpha}-\varepsilon_{\beta})/(\varepsilon_{\alpha}+\varepsilon_{\beta})$
is the dielectric contrast. ${\bf r}^{\star}= (x',y',-z')$ is the
location of the image of ${\bf r}'$ with respect to the
interface. The first term on the r.h.s of Eq. \ref{eq3} is the
direct Coulomb interaction and will generate the local Born
solvation energy upon integration over the charge distribution. The
last term in the first line of Eq. \ref{eq3} is the image charge
interaction, which can be either positive or negative depending on
whether the point charge is located on the high dielectric side or
low dielectric side; thus it either enhances or counteracts the
solvation energy effect.  
\begin{figure}[b]
\centering
\includegraphics[width=0.36\textwidth]{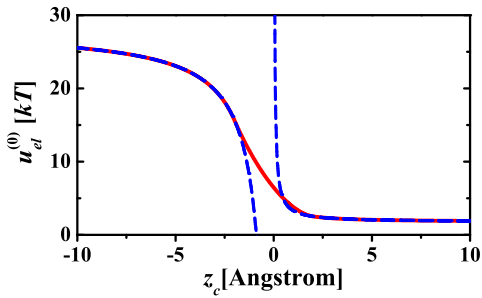}
\caption{Electrostatic self energy, $u_{el}^{(0)}$, of a monovalent
ion with uniform surface charge distribution, calculated by our
model (solid line) and the point charge model (dash line).
$\varepsilon_{1}=80$, $\varepsilon_{2}=5$ and $a=2\mathring{A}$.
 \label {1}}
\end{figure}

In the point-charge model $\rho ({\bf r}, {\bf r}_{c})=\nu_{\pm}
\delta ({\bf r}-{\bf r}_{c})$, Eq.~\ref{eq2} gives $u_{int}({\bf
r}_c)=2\pi l_B \nu_{\pm}^2 G({\bf r}_{c}, {\bf r}_{c})$, which
produces divergences in both the local Born solvation energy and in
the image charge interaction as $z_c \to 0$ from either side of the
interface. The use of a finite charge distribution avoids both types
of divergences.  Fig.~1 shows the result for the electrostatic self
energy, $u_{el}^{(0)}$, calculated for a nonpolarizable, uniform
surface charge distribution on the ion (thus
$u_{el}^{(0)}=u_{int}$). For comparison, we include the results from
the point-charge model, adjusted by the bulk Born energy
$\nu_{\pm}^2 l_B/2 a\varepsilon_{\alpha}$ on each side. While
$u_{el}^{(0)}$ calculated by the two models are consistent in the
bulk region ($\mid z_c \mid>a$), qualitative and dramatic
differences are seen in the interfacial region -- the most relevant
region for the interfacial activities of the ions.  Interestingly,
$u_{el}^{(0)}$ for an ion located exactly at the interface ($z_c=0$)
is significantly lower than the mean of the Born energy in two bulk
regions, reflecting the asymmetry in the image force between the two
dielectric media.

Polarization of the ion allows the charge distribution to
self-adjust to its local dielectric environment, which decreases
$u_{int}$ relative to that for a fixed uniform charge distribution.
However, this redistribution incurs an energy penalty $u_{pol}$.
Levin~\cite{Levin1} proposed a phenomenological model for $u_{pol}$
by taking reference to the perfectly conducting sphere and making a
Landau type of symmetry argument to describe this energy penalty. In
our notation $u_{pol}$ is:
\begin{equation}
u_{pol} ({\bf r}_c)= \frac{(\gamma_0 - \gamma)}{2 v \gamma} \int d
{\bf r} \left[\frac{\rho ({\bf r}, {\bf r}_{c})}{\rho_0}-1 \right]
^2 \label{eq4}
\end{equation}
where $\gamma$ is the polarizability of the ion, $\gamma_0 (=a^3)$
is the polarizability of a perfectly conducting sphere of the same
radius as the ion\cite{Jacksonbook}, $v$ is the volume of the ion,
and $\rho_0$ is the charge density for the uniform spherical
distribution on the ion surface.  The form of the coefficient in Eq.
\ref{eq4} was so constructed as to reproduce the known limiting
behavior, i.e., that it should be zero for the perfectly conducting
sphere and infinity for a nonpolarizable ion.

Putting together Eqs. \ref{eq2} and \ref{eq4}, we obtain the general
expression for $u_{el}$ with arbitrary charge distribution on the
ion.  The optimal distribution is then obtained from $\delta u_{el}
({\bf r}_c) / \delta \rho ({\bf r}, {\bf r}_{c})=0$. To avoid the
complexity of solving the high dimensional integral equation from
this condition, we make a variational trial function for $\rho ({\bf
r}, {\bf r}_{c})$.  We assume that polarization apportions
respectively $f$ and $1-f$ of the total ionic charge ($f \in [0,1]$)
uniformly to the two hemispheres of the ion separated by the $xy$
plane at $z_c$, i.e.,
\begin{equation}
\rho ({\bf r}, {\bf r}_{c})=\begin{cases}
2 f \rho_0({\bf r}) & for \; z \geq z_c\\
2 (1-f)\rho_0 ({\bf r}) & for \; z < z_c
\end{cases}
\label{eq6}
\end{equation}
where $\rho_0 ({\bf r}) = \nu_{\pm} \delta (\mid {\bf r}-{\bf r}_{c}
\mid -a)/4\pi a^2$ is the uniform surface distribution on the
sphere. The deviation of $f$ from $1/2$ measures the degree of
polarization of the ionic charge. Substituting the trial function
Eq. \ref{eq6} into Eqs. \ref{eq2} and \ref{eq4}, $u_{el}({\bf r}_c)$
can be simplified to a quadratic function of $f$, which can be
easily minimized to yield a position-dependent charge fraction
$f({\bf r}_{c})$. This optimal charge fraction $f({\bf r}_{c})$ is
then used to evaluate $u_{el}({\bf r}_c)$. Since the electrostatic
interaction includes the local Born solvation energy and the
long-range image force, the resulting polarization reflects the
combined effects of these terms. Fig.~2 shows the charge
polarization and the electrostatic self energy of I$^{-}$
($a_{I^-}=2.26\mathring{A}$, $\gamma_ {I^-}=6.9\mathring{A}^3$). In
the immediate vicinity of the interface ($\mid z_c \mid <a$),
I$^{-}$ is highly polarized. Charge polarization significantly lowers $u_{el}$
compared with the nonpolarizable ion.  Beyond
the immediate vicinity of the interface ($\mid z_c \mid>a$),
polarization is driven by the long-range image force, and $f$ decays
to $1/2$ as the ion approaches the bulk. While the effect of charge
polarization on $u_{el}$ is small on the high dielectric side beyond
$z_c =a$, $u_{el}$ of the polarizable ion is appreciably lower than
the nonpolarizable ion on the low dielectric side slightly beyond
$z_c =-a$, as a result of stronger and longer-range image force in
this region.

For comparison, in Fig.~2(b) we also include result from Levin's
polarizable ion model\cite{Levin1}. $u_{el}$ in Levin's theory only
extends to $z_c =-a$, whereas our theory yields a continuous
$u_{el}$ across the interface to the bulk air(oil) phase; this will
be important when there is appreciable ion partition in the oil
phase. In addition, $u_{el}$ on the low dielectric side is
significantly higher in our theory than from Levin's theory because
the relocation of charge from $z<0$ to $z>0$ changes the image force
from attractive to repulsive, which is not accounted for in Levin's
theory. The difference becomes more pronounced with increasing
dielectric contrast.
\begin{figure}[htb]
\centering
\includegraphics[width=0.4\textwidth]{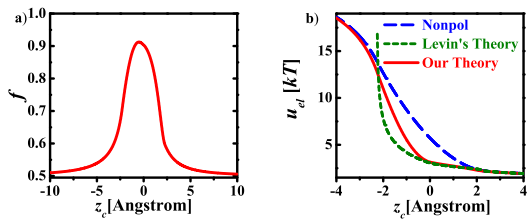}
\caption{(a) Charge polarization $f$, and (b) electrostatic self
energy $u_{el}$, for I$^{-}$.  For comparison, we include Levin's
theory\cite{Levin1} and a nonpolarizable ion of the same radius as
I$^{-}$ (dash line). $\varepsilon_{1}=80$, $\varepsilon_{2}=5$.
 \label {2}}
\end{figure}

The calculations so far concern only the electrostatic contributions
to the self energy, which will not be affected by the addition of
nonelectrostatic effects.  The total self energy of the ion is the
sum of the electrostatic and nonelectrostatic parts:
\begin{equation}
u({\bf r}_c)=u_{el}({\bf r}_c)+u_{ne}({\bf r}_c)\label{eq1}
\end{equation}
We now include the appropriate nonelectrostatic contributions to
discuss the interfacial behavior of different ions at the water/air
and water/oil interfaces respectively.

\textbf{Water/Air Interface} For the
nonelectrostatic self energy at the water/air interface, we take the
simplest form of cavity energy\cite{cavity1,Levin1}, which is the
work required to create a cavity for the ion where the water
molecules are excluded. It is given by\cite{cavity1,Levin1}
\begin{equation}
u_{ne}^{w/a} ({\bf r}_c)=\begin{cases}
\kappa a^3 & \; z_c \geq a\\
\frac{\kappa a^3}{4} \left(\frac{z_c}{a}+1 \right)^2
\left(2-\frac{z_c}{a}\right) & \; a > z_c \geq
-a\\
0   & \; z_c <
-a\\
\end{cases}
\label{eq5}
\end{equation}
with $\kappa \approx 0.3 \mathring{A} ^{-3}$ from bulk
simulation\cite{cavity2}.  $u_{ne}^{w/a}$ provides the driving
force for the ion to migrate from the bulk water to the interface;
this driving force is larger for larger ions. The self energy
profile of the ion across the interface is determined by the
competition between the cavity energy and the electrostatic self
energy, the former preferring the ion to reside on the air side and
the latter favoring it being on the aqueous side.
\begin{figure}[b]
\centering
\includegraphics[width=0.4\textwidth]{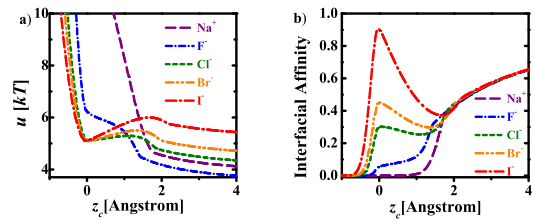}
\caption{(a) Self energy and (b) interfacial affinity of F$^-$,
Cl$^-$, Br$^-$, I$^-$ and Na$^+$ at the water/air interface.
$\varepsilon_{1}=80$, $\varepsilon_{2}=1$.
 \label {3}}
\end{figure}

Fig.~3(a) shows $u$ for four halogen anions and the alkali-metal
Na$^+$. We use the Born radius 2.26, 2.05, 1.91, 1.46 and 1.80
$\mathring{A}$\cite{Ionsize}, and the polarizability 6.90, 4.53,
3.50, 0.97 and 0.18 $\mathring{A}^3$\cite{Jungwirth}, respectively
for I$^-$, Br$^-$, Cl$^-$, F$^-$ and Na$^+$. For the larger and more
polarizable ions, such as I$^-$ and Br$^-$, the gain in cavity
energy at the relatively low cost of electrostatic self energy leads
to a local minimum at the interface in the self energy profile,
which is consistent with the result of MD simulation using a
polarizable potential model\cite{Dang}. For small and less
polarizable ions, such as F$^-$ and Na$^+$, $u$ is monotonic and
increases rapidly as the ion moves from the aqueous phase to air.
Our theory predicts a more repulsive $u_{el}$ for these ions on the
water side than in Ref.~\cite{Levin1}, which has the same
qualitative effect as the hydration effect considered in
Ref.~\cite{Levin2,Levin3} for predicting the surface tension.  For
more complex ions such as I${\rm O}_3^{-}$, explicit treatment of
hydration may be necessary\cite{Levin2,Levin3}.  However, the
quantitative importance of hydration needs to be reevaluated with
our more accurate electrostatic self energy.

The self energy of an ion is closely related to the concentration
profile of the ions.  While a full treatment has to include the
interaction between the ions, which leads to screening of the the
image forces, we can obtain a qualitative picture of the ion
distribution by defining the interfacial affinity as $e^{-[ u(z_c)-
u(\infty)]}$ to characterize the relative probability of finding the
ion in the interfacial region to the bulk.  In Fig.~3(b), we show
the interfacial affinity for the halogen anions and Na$^+$.  It is
clear that our theory captures the known specific ion effect, which
follows precisely the reverse Hofmeister series: I$^-
>$ Br$^->$ Cl$^->$ F$^-$\cite{Surfacetension,Hofmeister}. The local peak in the interfacial affinity of I$^-$ and
Br$^-$ ions is consistent with results of electron spectroscopy
experiments {\cite{Ghosal}} and computer simulations using
polarizable fields {\cite{Jungwirth}}. In addition, the interfacial
affinity of halogen anions is larger than that of Na$^+$, from which
we expect local charge separation and an induced electrical double
layer at the interface in a NaX solution, with the halogen anions
accumulating right around the location of the interface and the
Na$^+$ ions next to it on the water side. The electrostatic
potential gradient due to charge separation has been shown to be key
to explaining the Jones-Ray effect\cite{Onuki}.

\textbf{Water/Oil Interface} With a continuous self energy, our
theory naturally applies to the liquid/liquid interface.  In
addition to cavity energy, dispersion forces have been suggested to
be an important contribution to the nonelectrostatic self energy at
the water/oil interface\cite{Ninhamrev,dispersion,Levin3}. These
nonelectrostatic contributions set a chemical potential difference
between the two bulk phases in addition to the Born energy difference.
Phenomenologically, these nonelectrostatic effects can be captured
by a single parameter $B$ with a crossover in the interfacial
region that can be approximated by the interpolation scheme proposed
by Levin and coworkers\cite{Levin1,Levin2,Levin3}.
Similar to Eq. \ref{eq5}, we may write the nonelectrostatic self energy in the form:
\begin{equation}
u_{ne}^ {w/o} ({\bf r}_c)=\begin{cases}
B & \; z_c \geq a\\
\frac{B}{4} \left(\frac{z_c}{a}+1 \right)^2
\left(2-\frac{z_c}{a}\right) & \; a > z_c \geq
-a\\
0   & \; z_c <
-a\\
\end{cases}
\label{eq6}
\end{equation}

Restricting our consideration to cavity energy and dispersion force,
and taking the reference energy to be 0 in the bulk oil, $B=v^{w} -
v^{o} + A_{eff}(\gamma / \gamma_0)$\cite{Levin3}, where $v^{w}$ is
the cavity energy in water, which scales with the cavity volume for
small cavity sizes ($a<4\mathring{A}$) and with the surface area for
larger cavities\cite{cavity1}. $v^{o}$ is the cavity energy in oil,
which is primarily due to the surface energy between the ion and
oil\cite{Pierotti,Ninhamcavity}. $A_{eff}$ is the effective Hamaker
constant for the water/oil interface, estimated to be about $4
kT$\cite{Levin3} for a typical oil-water system.  Alternatively, we
can treat $B$ as an adjustable parameter from the bulk partitioning
of the ions between water and oil.

We defer a more general study of ions at the water/oil interface to
a future study.  Here we consider a special case of hydrophobic
ions. Schlossman and coworkers observed strong adsorption of
hydrophobic ions at the water/oil interface by X-ray
reflectivity\cite{hydrophobic}, from which it is inferred that there
is an attractive well for the self energy on the oil side. However,
no explanation has been given to the origin of this attractive well.
Within our theory, this phenomenon can be easily understood as
arising from the long-range image charge attraction of the
hydrophobic ions in the low-dielectric oil phase. Fig. 4 shows the
self energy and interfacial affinity of a hydrophobic ion calculated
by our theory with $B=33 kT$, estimated using $A_{eff}= 4 kT$,
surface tension of water and surface tension of the oil used in the
experiment\cite{paraB}. As the ion approaches the interface from the
oil side, $u$ decreases because of the image charge attraction, and
then increases rapidly due to the unfavorable contact with the
aqueous environment. The peak in the interfacial activity on the oil
side of the interface corresponds to minimum in the self energy with
depth of about $5.8kT$, in good agreement with the experimental
results\cite{hydrophobic}. We note that although the choice of $B$ will affect bulk partitioning of the ions,
the depth of the attractive well is quite insensitive to the precise numerical value
as long as $B$ is large enough to ensure
hydrophobicity of the ion.  This clearly demonstrates the electrostatic origin of the strong
adsorption of hydrophobic ions on the oil side of the interface, as the nonelectrostatic contributions (as
depicted by the dash line) do not contain an attractive well.

\begin{figure}[htbp]
\centering
\includegraphics[width=0.4\textwidth]{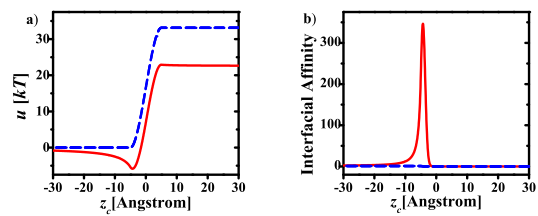}
 \caption{(a) Self energy and (b) interfacial affinity of a hydrophobic ion at the water/oil
interface relative to the bulk oil. The dash line shows the results
calculated by the nonelectrostatic contribution alone.
$a=5\mathring{A}$, $\gamma / \gamma_0=0.5$, $\varepsilon_{1}=80$ and
$\varepsilon_{2}=5$. $B=33 kT$\cite{paraB}.
 \label {5}}
\end{figure}

In conclusion, by treating both the short-range (solvation) and
long-range (image force) electrostatic forces as well as charge
polarization induced by these forces in a consistent manner, we
obtain a simple continuous electrostatic self energy across the
interface, making it applicable to both water/air and liquid-liquid
interfaces.  A systematic and accurate treatment of the
electrostatic self energy is essential for evaluating the relative importance of the
nonelectrostatic contributions.  Combining the electrostatic self
energy with existing models for nonelectrostatic contributions, we are able to explain a
number of interfacial specific ion effects using the intrinsic
parameters of the ion, such as the valency, radius, and
polarizability.  The self energy model developed here provides the
essential ingredient in a complete theory to treat ions at finite
concentration, via e.g., the weak coupling
theory\cite{weak-coupling} or modified Poisson-Boltzmann
theory\cite{Netz}, to describe the phenomena mentioned in
the beginning of this Letter.

\begin{acknowledgments}
Acknowledgment is made to the Donors of the American Chemical
Society Petroleum Research Fund for partial support of this
research. We thank Prof. Yan Levin for many helpful discussions.
\end{acknowledgments}


\begin{thebibliography}{}
\bibitem{paraB}
For this large ion, the cavity energy is proportional to its surface
area\cite{cavity1}: $v^w - v^o = 4 \pi a^2 (\sigma^{iw}
-\sigma^{io})$, where $\sigma^{iw}$ and $\sigma^{io}$ are
respectively the ion-water and ion-oil interfacial energy.  We
estimate $\sigma^{iw}$ and $\sigma^{io}$ in two ways.  Taking the
ion as a cavity (vacuum), these two surface energies can be
approximated by the surface tension of water $\sigma^w=72.8mN/m$ and
the oil (1,2-Dichloroethane \cite{hydrophobic})
$\sigma^o=38.75mN/m$. This yields $B \approx 33kT$.  If we take the
hydrophobic ion to be of the same material as the oil, then
$\sigma^{io}=0$. $\sigma^{iw}=\sigma^{ow}$, where
$\sigma^{ow}=28.2mN/m$ is the interfacial tension between water and
1,2-Dichloroethane\cite{hydrophobic}. This estimate gives $B \approx
28kT$.
\end{thebibliography}
\end{document}